\documentclass[12pt]{article}
\usepackage{eurosym}
\usepackage{graphicx}
\usepackage[utf8]{inputenc}
\usepackage[T1]{fontenc}
\usepackage{indentfirst}
\usepackage[margin=0.8in,nomarginpar]{geometry}
\usepackage[final]{hyperref}
\usepackage{amsmath}
\usepackage{hyperref}
\usepackage{cite}
\usepackage{subcaption}
\usepackage{caption}
\usepackage{amssymb}
\usepackage{multirow}
\usepackage{graphicx}
\usepackage{color}
\usepackage{mathtools, nccmath}
\usepackage{amssymb, amsthm, mathrsfs}

\hypersetup{
colorlinks=true,
linkcolor=blue,
citecolor=blue,
filecolor=magenta,
urlcolor=blue
}

\begin{document}

\title{The Condensation of Ideal Dunkl-Bose Gas in Power-Law Traps}
\author{A. Hocine\thanks{%
a.hocine@univ-chlef.dz} \\
Laboratory for Theoretical Physics and Material Physics Faculty of Exact \\
Sciences and Informatics, Hassiba Benbouali University of Chlef, Algeria \
\and F. Merabtine\thanks{%
merabtinefateh172@gmail.com} \\
Laboratory for Theoretical Physics and Material Physics Faculty of Exact \\
Sciences and Informatics, Hassiba Benbouali University of Chlef, Algeria \ \
\ \ \thinspace  \and B. Hamil\thanks{%
hamilbilel@gmail.com} \\
D\'{e}partement de physique, Facult\'{e} des Sciences Exactes, \\
Universit\'{e} Constantine 1, Constantine, Algeria. \and B. C. L\"{u}tf\"{u}o%
\u{g}lu\thanks{%
bekir.lutfuoglu@uhk.cz (Corresponding author)} \\
Department of Physics, University of Hradec Kr\'{a}lov\'{e},\\
Rokitansk\'{e}ho 62, 500 03 Hradec Kr\'{a}lov\'{e}, Czechia.\ \  \and M.
Benarous\thanks{%
m.benarous@univ-chlef.dz} \\
Laboratory for Theoretical Physics and Material Physics Faculty of Exact \\
Sciences and Informatics, Hassiba Benbouali University of Chlef, Algeria \ \
\ }
\date{}
\maketitle

\begin{abstract}
We explore the phenomenon of Bose-Einstein condensation in two and one-dimensional Dunkl-boson gases confined within a power-law potential, employing the framework of Dunkl-deformed boson theory. Our investigation involves the calculation of particle numbers and phase transition temperatures using the Dunkl formalism. To assess the validity of our findings, we compare them with the corresponding results obtained from the standard approach. We find that the impact of the Dunkl-formalism on the condensate fractions is similar in one and two-dimensional cases. However, we see that this conclusion is not fully valid for the phase transition temperature.  
\end{abstract}

\section{Introduction}

In the literature, particles with integer and half-integer spin are classified as bosons and fermions, respectively. Several properties of the ideal Bose gas, consisting only of bosons, differ substantially from the ideal Fermi gas, involving only fermions. For example, at sufficiently low temperatures bosons predominantly occupy the ground state with zero energy and momentum while fermions do not. Furthermore, below a critical temperature, the ground state occupancy number tends to be nearly equal to the total number of the bosons of the system, and the particles' characteristic de Broglie wavelength becomes macroscopically large signifying the overlap of the bosons \cite{lewis, landu, van der berg}. This quantum phase transition, which is known as Bose-Einstein condensation (BEC),  was initially predicted theoretically by Bose \cite{Bose} and Einstein \cite{Einstin, Ein2} in the mid-1920s. Since then, several other phenomena have been identified as manifestations of the BEC, such as superfluidity in liquid helium \cite{London, Glyde, Zhu2015}, high-temperature superconductivity in certain materials \cite{Rosencwaig, Hasimoto, Cheng}, BEC of hypothetical Higgs particles \cite{Sannino, Han, Phan} and pions\cite{Bialas, Begun2008, Deb}, among others. 

After many attempts, the first observation of BEC in dilute gases was achieved in 1995.  Two groups of the Joint Institute for Laboratory Astrophysics (JILA) cooled down vapors of magnetically trapped rubidium and sodium atoms to very low temperatures and detected the quantum phase transitions over a limited time interval \cite{anderson, Davis}. The same year, Bradley et al showed the presence of BEC for the confined lithium atoms \cite{Bradley}.  For those achievements, Cornell, Wierman, and Kettle shared the Nobel Prize in Physics in 2001 \cite{Nobel}. With this motivation, studies in the field of BEC of ultracold quantum gases progressed by considering various atoms and trapping interactions in different geometries at different low temperatures \cite{7, 8, 9, 11, 12, 13}. For example, in some experimental works researchers focused on creating quasi-two-dimensional and quasi-one-dimensional trapped gases by tightly confining the particles' motion in one or two directions \cite{Gorlitz, Burger, Schreck, Greiner}. In general, the confinement force determines whether the gas system is two-dimensional or one-dimensional, depending on the interparticle interaction value between the particles. In other words, shallow confinement through the other direction or directions allows for describing a trapped one or two-dimensional gas. In most cases, the confining traps are well approximated by harmonic potentials \cite{gro1, gro2, kirsten1, kirsten2, napo, zeng1, zeng2}. In the literature, generic power-law potential energy is also employed as a trapping potential energy for the examination of the BEC phenomena. For example, in an intriguing study, Bagnato et al conducted research on an ideal Boson gas in three dimensions to explore the critical temperature and ground state population by utilizing a generic power-law potential energy form \cite{Bagnato 1}. In another study, they investigated the BEC phase transition in low-dimensional traps with power-law potentials they suggested an experimental realization of a two-dimensional Bose gas \cite{Bagnato 2}. For further information, we refer readers to see the review articles  \cite{dalefo, corn, Bloch, Lahaye} and the references therein. In the last two decades, this intriguing quantum phenomenon still has gathered attention and it is addressed by studies in different disciplines \cite{Jao, Jao1, Anglin, Shem, Aftergood, Kasprzak, Griesmaier, Aikawa, Sun, Santos, Lekala, Chavez, Takeuchi, Cenatiempo, Ferreira, Keeling, Levkov}.


On the other hand, half a century ago, in ongoing research in a different field of science, pure mathematicians were investigating the relationship between reflection groups and differential-difference operators. In 1989, Dunkl defined a self-adjoint operator,  which later became known as the Dunkl derivative 
\begin{eqnarray}
D_i=\frac{\partial}{\partial x_i}+\frac{\theta}{x_i}\left(1-R_i\right),  \label{Dunk1}
\end{eqnarray}
to examine the kernel solutions of the corresponding Laplacian that are commonly referred to as h-harmonic functions \cite{Dunkl1}. In fact, the Dunkl operator looks similar to the Yang operator, which was defined by Yang in the mid-twentieth century \cite{Yang} after Wigner's pioneering work discussing the determination of commutation relations from equations of motion \cite{Wigner}. Therefore, in the literature, $\theta$ is preferred to be named as the Wigner constant. Here, $R_i$ corresponds to the reflection operator in any direction which obeys the following relations
\begin{align}
R_if(x_i)=f(-x_i),&\qquad R_iR_j=R_jR_i,\qquad R_ix_j=-\delta_{ij}x_jR_i,
\notag \\
&\qquad R_i\frac{\partial}{\partial x_i}=-\delta_{ij}\frac{\partial}{%
\partial x_j}R_i.
\end{align}
In addition to its wide usage in the field of pure mathematics \cite{Dunkl2, Rosler}, the Dunkl operator also appeared in the field of theoretical physics, i.e. in the solutions of quantum Calogero-Moser-Sutherland models \cite{Lapointe, Kakei, Chakra} and deformed quantum mechanics \cite{Mik1, Gamboa, Mik2, Klishevich, Horvathy}. The reflection operator inside the Dunkl operator creates different impacts depending on whether the function it affects is even or odd. From this point of view, especially in the last decade, the Dunkl operator has begun to be used instead of the ordinary partial derivative operator in the differential equations of relativistic and non-relativistic quantum mechanical systems. For example in \cite{G1, G2, G3, G4}, Genest et al investigated the relativistic and non-relativistic Dunkl-isotropic and Dunkl-anisotropic harmonic oscillator solutions with their symmetries in two and three dimensions. Following these non-relativistic regime studies, the analytical solutions of the relativistic regime oscillators, namely Dunkl-Dirac and Dunkl-Klein-Gordon oscillators, are given for one, two, and three dimensions in \cite{Sargol, Mota1, Bilel3, Mota2, Mota3, Bilel1}. In addition, solutions of Duffin-Kemmer-Petiau and Pauli equations are derived within the Dunkl formalism, respectively in \cite{DunklDKP, DunklPauli}. Nowadays, we observe that Dunkl formalism is being adopted in many distinct sub-fields of physics, i.e.  Newtonian mechanics \cite{Chungrev}, electromagnetic theory \cite{ChungDunkl},  statistical mechanics \cite{Marcelo}, solid-state physics \cite{BilelGr}, general relativity \cite{DRelat1}, and black hole thermodynamics \cite{DRelat2}.

Very recently in a series of articles, we have discussed ideal Bose gas systems and their condensates using the Dunkl formalism. More precisely, in \cite{Mer}, we considered an ideal Bose gas system and demonstrated the impact of the Dunkl formalism on its critical temperature. Then, in \cite{BilelStat}, we studied the Dunkl-BEC in the presence of the gravitational field for two and three-dimensional cases. Next in \cite{BCLarxiv}, we considered isotropic harmonic oscillator potential energy as the trapping potential energy and examined the Dunkl-BEC with the Dunkl-deformed thermal quantities. In this fourth consecutive work, we handle the same problem but with another important trapping potential energy, namely the power-law potential energy. Our main goal is to determine the impact of the Dunkl formalism on the condensate for one and two-dimensional cases. To this end, we construct the manuscript as follows: In section 2, we consider the one-dimensional power-law trap to derive the Dunkl-BEC temperature and condensate fraction. Then, in section 3, we handle the two-dimensional power-law trap and derive the Dunkl-BEC temperature and condensate fraction. After comparing the latter findings with the one-dimensional results, we briefly conclude the manuscript in the last section.

\section{Dunkl-BEC in one-dimensional power-law potential trap}

Let us consider a nonrelativistic ideal atomic gas trapped in a
one-dimensional power-law potential with the energy spectrum
\begin{eqnarray}
\epsilon \left( p,x\right) =\frac{p^{2}}{2m}+U_{0}\left( \frac{|x|}{L}%
\right) ^{\eta },   \label{e1}
\end{eqnarray}%
where $p$ and $m$ correspond to the momentum and mass of identical gas atoms. Here, $U_{0}$, $L$, and $\eta $, where ($\eta <2$),  are positive constants that describe the external potential. Now, let us derive the density of states via the semiclassical formula 
\begin{eqnarray}
\rho (\epsilon )=\int \int \frac{dxdp}{\sqrt{2\pi \hbar }}\delta \left(\epsilon -\epsilon \left( p,x\right) \right) .
\end{eqnarray}%
After performing the simple algebra we find the density of states function as
\begin{eqnarray}
\rho (\epsilon )=\frac{2L\sqrt{2m}}{h}\frac{\epsilon ^{\frac{1}{\eta }-\frac{%
1}{2}}}{\eta U_{0}^{\frac{1}{\eta }}}F(\eta ),  \label{de}
\end{eqnarray}%
where
\begin{eqnarray}
F(\eta )=\int_{0}^{1}\frac{y^{\frac{1}{\eta }-1}}{\sqrt{1-y}}dy=\frac{\sqrt{%
\pi }\Gamma (1/\eta )}{\Gamma \left( 1/\eta +1/2\right) }.
\end{eqnarray}%
Here,  $\Gamma \left( l\right) =\int_{0}^{\infty }e^{-x}x^{l-1}dx$, is the gamma function.

Usually in the literature, the BEC is described within the grand canonical ensemble. In this context the number of particles in thermodynamical equilibrium reads:
\begin{eqnarray}
N=N_{0}+N_{e},
\end{eqnarray}%
where $N_{0}$ and $N_{e}$ stand for the average number of the ground state and thermal (excited states) particles, respectively.

Now, we take into account the Dunkl formalism. In this case, the number of condensed and thermal particles differ from the conventional ones by extra Dunkl-correction terms \cite{Mer}. (For technical details see the third section of \cite{Mer}.)
\begin{eqnarray}
N_{0}^{D}&=&\frac{2}{z^{-2}-1}+\frac{1+2\theta }{z^{-(1+2\theta )}+1}, \\
N_{e}^{D}&=&\sum_{i\neq 0}^{\infty }\frac{2}{e^{2\beta \epsilon _{i}}z^{-2}-1}+%
\frac{1+2\theta }{e^{\beta (1+2\theta )\epsilon _{i}}z^{-(1+2\theta )}+1}.
\end{eqnarray}
Here, $\beta$ is the thermodynamic temperature, $\beta =\frac{1}{KT}$; $z$ is the fugacity,  $z=e^{\beta \mu }$; and $\mu $ is the chemical potential.

In a semi-classical approach, one can use the density of state function as a weight function and substitute the sum with the integral by assuming the energy spectrum is continuous. Then, the number of particles reads:
\begin{eqnarray}
N=N_{0}^{D}+\int_{0}^{\infty }d\epsilon \rho (\epsilon )\left[ \frac{2}{%
e^{2\beta \epsilon }z^{-2}-1}+\frac{1+2\theta }{e^{\beta (1+2\theta
)\epsilon }z^{-(1+2\theta )}+1}\right] .  \label{nn}
\end{eqnarray}%
Substituting Eq. (\ref{de}) into Eq. (\ref{nn}), we obtain
\begin{eqnarray}
N=N_{0}^{D}+\frac{2L\sqrt{2m}}{h}\frac{F(\eta )}{\eta U_{0}^{\frac{1}{\eta }}%
}\left( KT\right) ^{\frac{1}{\eta }+\frac{1}{2}}\Gamma \left( \frac{1}{\eta }%
+\frac{1}{2}\right) g_{\frac{1}{2}+\frac{1}{\eta }}(z,\theta ),  \label{4}
\end{eqnarray}%
where%
\begin{eqnarray}
g_{\frac{1}{2}+\frac{1}{\eta }}(z,\theta )=g_{\frac{1}{2}+\frac{1}{\eta }%
}(z)+g_{\frac{1}{2}+\frac{1}{\eta }}(-z)-\left( 1+2\theta \right) ^{\frac{1}{%
2}-\frac{1}{\eta }}g_{\frac{1}{2}+\frac{1}{\eta }}\left( -z^{1+2\theta
}\right), \label{db}
\end{eqnarray}
is the generalized Dunkl-Bose function \cite{BCLarxiv}. It is worth noting that in the absence of the Wigner parameter, in other words for $\theta=0$,  Eq. (\ref{db}) becomes the same as of the ordinary Bose function \cite{Pathria}. In this case, the total number of particles reads:
\begin{eqnarray}
N=N_{0}+\frac{2}{\eta }\frac{\sqrt{2m}}{h}L\frac{F(\eta )}{U_{0}^{\frac{1}{%
\eta }}}\left( KT\right) ^{\frac{1}{\eta }+\frac{1}{2}}\Gamma \left( \frac{1%
}{2}+\frac{1}{\eta }\right) g_{\frac{1}{2}+\frac{1}{\eta }}(z).  \label{25}
\end{eqnarray}
When the temperature of the system drops to the critical temperature, phase transition initiates, and the number of excited particles becomes (nearly) equal to the total number of particles, $N_{0}^{D}= 0$, $z=1$. Therefore, Eq. \eqref{4} gives
\begin{eqnarray}
N=\frac{2}{\eta }\frac{\sqrt{2m}}{h}L\frac{F(\eta )}{U_{0}^{\frac{1}{\eta }}}%
\left( KT_{c}^{D}\right) ^{\frac{1}{\eta }+\frac{1}{2}}\Gamma \left( \frac{1%
}{2}+\frac{1}{\eta }\right) g_{\frac{1}{2}+\frac{1}{\eta }}(1).  \label{26}
\end{eqnarray}
Thus, we can express the Dunkl-BEC temperature of ideal Bose gases in a one-dimensional external power-law trap as
\begin{eqnarray}
T_{c}^{D} =\frac{1}{K} \Bigg[\frac{\eta }{2}\frac{Nh}{\sqrt{2m}L}\frac{%
U_{0}^{\frac{1}{\eta }}}{F(\eta )}\frac{1}{\Gamma \left( \frac{1}{2}+\frac{1%
}{\eta }\right) g_{\frac{1}{2}+\frac{1}{\eta }}(1,\theta )}\Bigg]^{\frac{%
2\eta }{\eta +2}}.  \label{eqn}
\end{eqnarray}
Here, we can re-express the Dunkl-Bose function in terms of $\Gamma $ and $\text{PolyLog}$ functions as%
\begin{eqnarray}
g_{\frac{1}{2}+\frac{1}{\eta }}(1,\theta )=\frac{\text{PolyLog}\left( \frac{1%
}{2}+\frac{1}{\eta },1\right) }{\Gamma \left( \frac{1}{2}+\frac{1}{\eta }\right) }\left\{ 1+\left[ 1-\left( 1+2\theta \right) ^{\frac{1}{2}-\frac{1}{\eta }}\right] \frac{\text{PolyLog}\left( \frac{1}{2}+\frac{1}{\eta }%
,-1\right) }{\text{PolyLog}\left( \frac{1}{2}+\frac{1}{\eta },1\right) }%
\right\} .
\end{eqnarray}%
Properties of the $\text{PolyLog}$ function show that  $g_{\frac{1}{2}+\frac{1}{\eta }}(1,\theta )$ can be finite if and only if the condition $\eta <2$ exists. Thus, we conclude that the Dunkl-BEC can occur only for $\eta <2$ values of the power-law potentials. 

\begin{figure}[htb!]
\begin{minipage}[t]{0.5\textwidth}
        \centering
        \includegraphics[width=\textwidth]{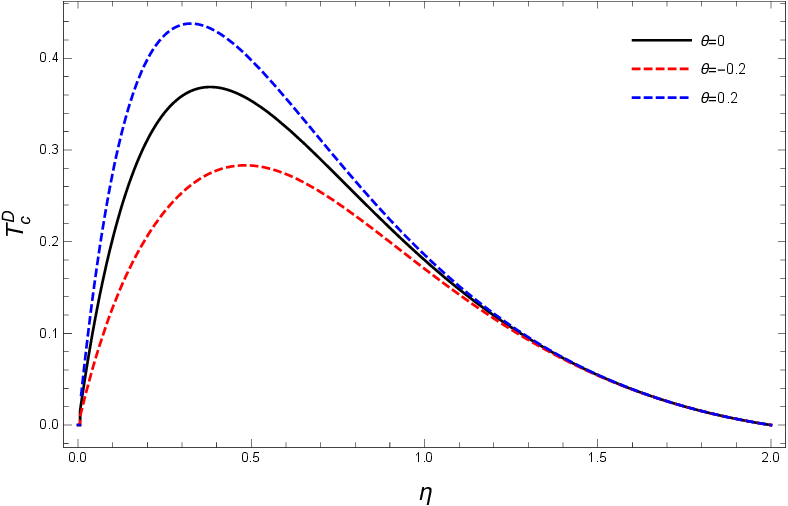}

       \subcaption{}\label{fig:pcT1a}
   \end{minipage}%
\begin{minipage}[t]{0.5\textwidth}
        \centering
        \includegraphics[width=\textwidth]{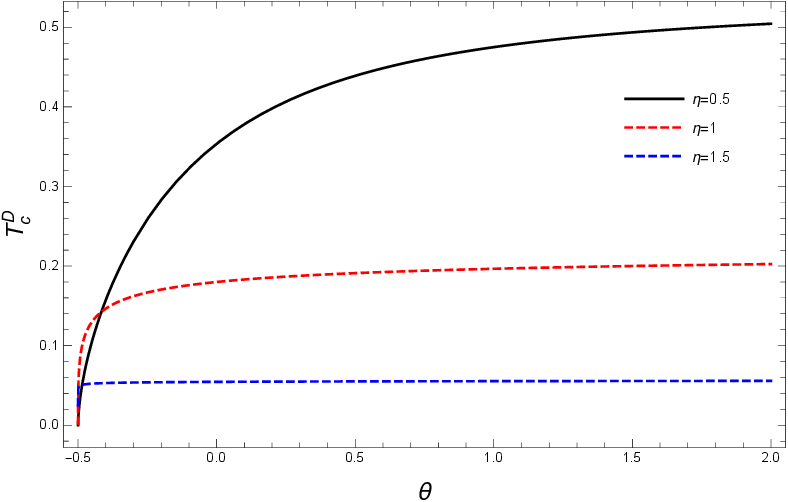}
        \subcaption{}\label{fig:pcT1b}
   \end{minipage}
\caption{Characteristic behavior of the Dunkl-corrected critical temperature function of the one-dimensional trapped system for $U_0=m=h=K=N=1$.}
\label{nfig1}
\end{figure}

In Figure \ref{nfig1}, we qualitatively examine the characteristic behavior of the Dunkl-BEC temperature. To this end, in Figure \ref{fig:pcT1a}, we depict its variation versus the potential coefficient for three different values of the Wigner parameter. We see that the critical temperature has a peak value for the power-law potential parameter between $0$ and $2$. We find that the critical temperature, and hence the peak value, varies depending on the Wigner parameter. More precisely, we observe that it takes smaller values for negative Wigner parameters, while it takes larger values for positive ones. Then, in Figure \ref{fig:pcT1b} we plot the Dunkl-BEC temperature variation versus the Wigner parameter for three different values of the potential coefficient. We observe that for greater values of $\eta$ the effect of the Dunkl formalism appears relatively smaller than the others.

\begin{figure}[htb!]
\begin{minipage}[t]{0.5\textwidth}
        \centering
        \includegraphics[width=\textwidth]{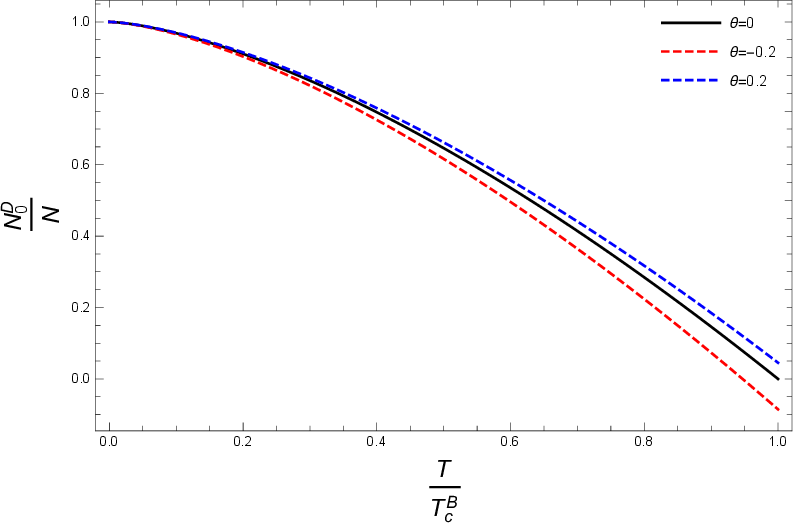}

       \subcaption{ $ \eta=1$}\label{fig:pc1}
   \end{minipage}%
\begin{minipage}[t]{0.5\textwidth}
        \centering
        \includegraphics[width=\textwidth]{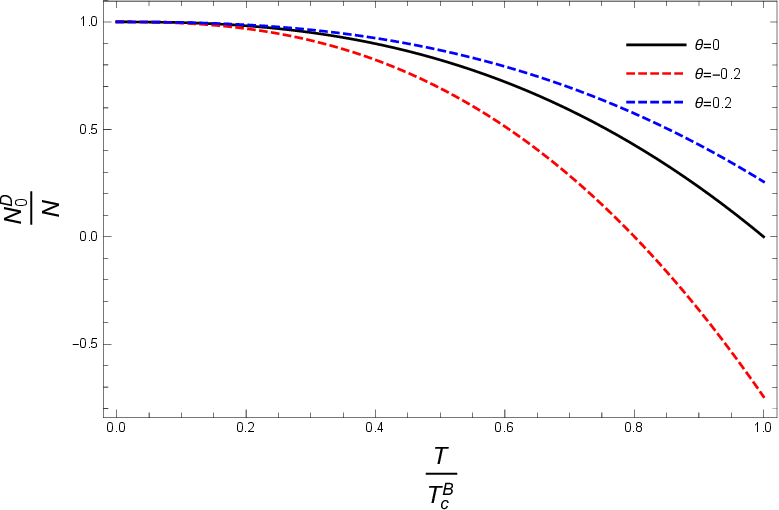}
         \subcaption{ $\eta=0.5$.}\label{fig:pd1}
   \end{minipage}
\caption{The ground state population versus $\frac{T}{T_{c}^{D}}$
for different values of the Wigner parameter.}
\label{nfig2}
\end{figure}

\newpage
It is worth mentioning that in the limit of $\theta \rightarrow 0,$ Eq. (\ref{eqn}) reduces to the
standard result%
\begin{eqnarray}
T_{c}^{B} =\frac{1}{K} \Bigg[\frac{\eta }{2}\frac{Nh}{\sqrt{2m}}\frac{%
U_{0}^{\frac{1}{\eta }}}{F(\eta )}\frac{1}{\Gamma \left( \frac{1}{2}+\frac{1%
}{\eta }\right) g_{\frac{1}{2}+\frac{1}{\eta }}(1)}\Bigg]^{\frac{2\eta }{%
\eta +2}}.
\end{eqnarray}%
Then, we relate the Dunkl-modified critical temperature to the standard critical temperature. $T_{c}^{B}$, as 
\begin{eqnarray}
\frac{T_{c}^{D}}{T_{c}^{B}}={\left( 1+\frac{g_{\frac{1}{2}+\frac{1}{%
\eta }}(-1)}{g_{\frac{1}{2}+\frac{1}{\eta }}(1)}\left[ 1-\left( 1+2\theta
\right) ^{\frac{1}{2}-\frac{1}{\eta }}\right] \right) ^{-\frac{2\eta }{\eta +2%
}}}.
\end{eqnarray}
Next, we examine the condensate fraction. To this end, we express the particle ratio in the ground state, as a function of the temperature. With the help of Eqs. (\ref{25}) and (\ref{26}), we get the fraction
\begin{eqnarray}
\frac{N_{0}^{D}}{N}=1-\left( \frac{T}{T_{c}^{D}}\right) ^{\frac{1}{\eta }+%
\frac{1}{2}}.
\end{eqnarray}
for $T<T_{c}^{D}$. Then, we re-express it as
\begin{eqnarray}
\frac{N_{0}^D}{N}=1-\left( 1+\frac{g_{\frac{1}{2}+\frac{1}{\eta }}(-1)}{g_{%
\frac{1}{2}+\frac{1}{\eta }}(1)}\left[ 1-\left( 1+2\theta \right) ^{\frac{1}{%
2}-\frac{1}{\eta }}\right] \right) \left( \frac{T}{T_{c}^{B}}\right) ^{\frac{%
1}{\eta }+\frac{1}{2}}.
\end{eqnarray}%
We note that for the absence of Dunkl formalism, this ratio reduces to
\begin{eqnarray}
\frac{N_{0}}{N}=1-\left( \frac{T}{T_{c}^{B}}\right) ^{\frac{%
1}{\eta }+\frac{1}{2}}.
\end{eqnarray}%
In Figure \ref{nfig2}, we plot the condensate fraction versus the normalized temperature. 

\newpage
We observe that the ground state population is a monotonically decreasing function. We observe that the Wagner parameter increases or decreases the tendency of the ground state population function to decrease, depending on its positive or negative value. This effect increases as the temperature approaches its critical value.

\section{Dunkl-BEC in two-dimensional power-law potential trap}

\begin{figure}[htb!]
\begin{minipage}[t]{0.5\textwidth}
        \centering
        \includegraphics[width=\textwidth]{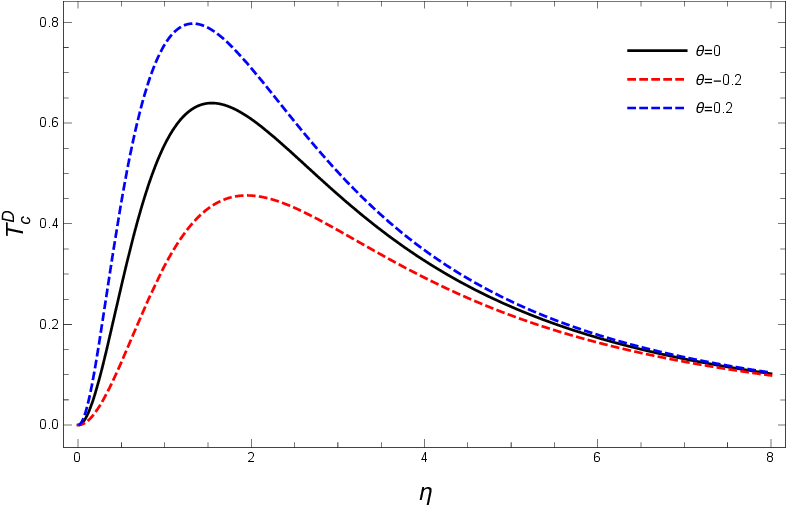}

       \subcaption{}\label{fig:pcT2a}
   \end{minipage}%
\begin{minipage}[t]{0.5\textwidth}
        \centering
        \includegraphics[width=\textwidth]{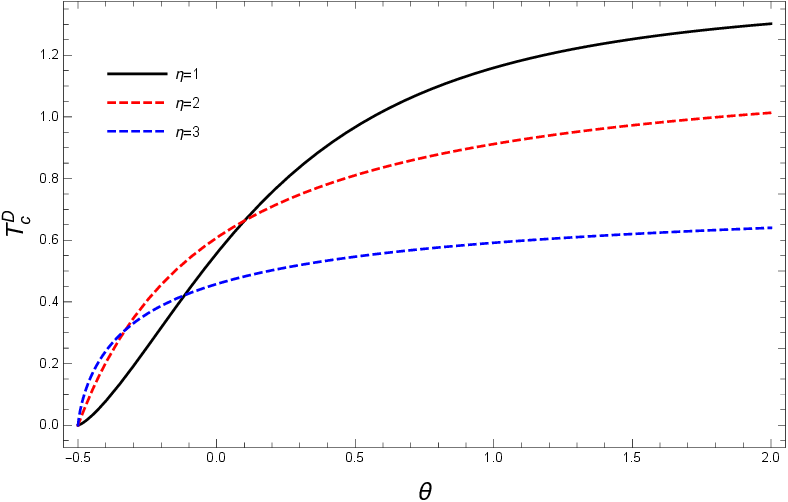}
        \subcaption{}\label{fig:pcT2b}
   \end{minipage}
\caption{Characteristic behavior of the Dunkl-corrected critical temperature function of the two-dimensional trapped system for $U_0=m=h=K=N=1$.}
\label{nfig3}
\end{figure}

In this section, we consider a two-dimensional power-law trapping potential and we examine the Dunkl-BEC phenomena in a similar way that we have already performed above. In the latter case, the density of states of a two-dimensional gas confined by the trapping potential $U(r)=U_{0}\left( \frac{r}{a}\right) ^{\eta }$ reads: 
\begin{eqnarray}
\rho (\epsilon )=\frac{2\pi ^{2}ma^{2}}{h^{2}}\left[ \frac{\epsilon }{U_{0}}%
\right] ^{\frac{2}{\eta }}.  \label{dnn}
\end{eqnarray}
First, we substitute the latter density of states expression into Eq. \eqref{nn}, and we get the total number
\begin{eqnarray}
N=N_{0}^{D}+\frac{2\pi ^{2}ma^{2}}{h^{2}}\int_{0}^{\infty }d\epsilon )\left[
\frac{2\epsilon ^{\frac{2}{\eta }}}{e^{2\beta \epsilon }z^{-2}-1}+\frac{%
\left( 1+2\theta \right) \epsilon ^{\frac{2}{\eta }}}{e^{\beta (1+2\theta
)\epsilon }z^{-(1+2\theta )}+1}\right] ,  \label{dnn1}
\end{eqnarray}%
then, we rewrite it in  the following form 
\begin{eqnarray}
N=N_{0}^{D}+\frac{2\pi ^{2}ma^{2}}{h^{2}}\frac{1}{U_{0}^{\frac{2}{\eta }}}%
\left( KT\right) ^{\frac{2}{\eta }+1}\Gamma \left( \frac{2}{\eta }+1\right)
g_{\frac{2}{\eta }+1}(z,\theta ),  \label{11}
\end{eqnarray}%
with the terms of the Dunkl-Bose function
\begin{eqnarray}
g_{\frac{2}{\eta }+1}(z,\theta )=g_{\frac{2}{\eta }+1}(z)+g_{\frac{2}{\eta }%
+1}(-z)-\left( 1+2\theta \right) ^{-\frac{2}{\eta }}g_{\frac{2}{\eta }%
+1}(-z^{1+2\theta }).
\end{eqnarray}%
Next, we assume that the temperature is equal to the critical temperature, $T=T_{c}^{D}$, thus, $N_{0}^{D}=0$ and $z=1$. So that we obtain the Dunkl-BEC temperature 
\begin{eqnarray}
 T_{c}^{D} =\frac{1}{K}\left[ \frac{Nh^{2}U_{0}^{\frac{2}{\eta }}}{2\pi
^{2}ma^{2}}\frac{1}{\Gamma \left( \frac{2}{\eta }+1\right) g_{\frac{2}{\eta }%
+1}(1,\theta )}\right] ^{\frac{\eta }{2+\eta }}.  \label{266}
\end{eqnarray}%
In Figure \ref{nfig3}, We qualitatively demonstrate the characteristic behavior of the Dunkl-BEC temperature. More precisely, we present the variation of $T_{c}^{D}$ for the potential parameter for different values of the Wigner parameter in Figure \ref{fig:pcT2a}. We observe that the behavior in two dimensions is the same behavior of the one-dimensional, i.e. the critical temperature has a peak value of the critical temperature and it decreases or increases depending on the value of the Wigner parameter. However, there are also differences. In the two-dimensional power-law trap, there is no constraint on the $\eta$ parameter. Therefore, we conclude that the Dunkl-BEC can occur for any positive and finite values of $\eta $ in principle. We show the variation of the $T_{c}^{D}$ versus the Wigner parameter in Figure \ref{fig:pcT2b} for three values of the potential parameter.  We observe that the impact of the Wigner parameter is different when it takes negative or positive values. This chaotic effect also shows different characteristics for different potential parameter values. 

It is also worth noting that as the Wigner parameter approaches zero, the Dunkl-critical temperature converges to the conventional Bose critical temperature, $T_{c}^{B}$, given in the form of
\begin{eqnarray}
T_{c}^{B} =\frac{1}{K}\left[ \frac{Nh^{2}U_{0}^{\frac{2}{\eta }}}{2\pi^{2}ma^{2}}\frac{1}{\Gamma \left( \frac{2}{\eta }+1\right) g_{\frac{2}{\eta }+1}(1)}\right] ^{\frac{\eta }{2+\eta }}.  \label{277}
\end{eqnarray}%
By comparing Eqs. (\ref{266}) and (\ref{277}), we determine a relationship between $T_{c}^{D}$ and $T_{c}^{B}$ as 
\begin{eqnarray}
\frac{T_{c}^{D}}{T_{c}^{B}}=\left( {1+\frac{g_{\frac{2}{\eta }+1}(-1)}{g_{\frac{2}{\eta }+1}(1)}\left[ 1-\left( 1+2\theta \right) ^{-\frac{2}{\eta }}\right] }\right) ^{-\frac{\eta }{2+\eta }}.
\end{eqnarray}%

Finally, we consider the temperature $T\leq T_{c}^{D}$. In this case, we find the ground state population $\frac{N_{0}^{D}}{N}$ in the following form:
\begin{eqnarray}
\frac{N_{0}^{D}}{N}=1-\left( 1+\frac{g_{\frac{2}{\eta }+1}(-1)}{g_{\frac{2}{\eta }+1}(1)}\left[ 1-\left( 1+2\theta \right) ^{-\frac{2}{\eta }}\right]\right) \left( \frac{T}{T_{c}^{B}}\right) ^{\frac{2}{\eta }+1}. 
\end{eqnarray}
In the absence of the Dunkl formalism, the condensate fraction reduces to 
\begin{eqnarray}
\frac{N_{0}}{N}=1- \left( \frac{T}{T_{c}^{B}}\right) ^{\frac{2}{\eta }+1}. 
\end{eqnarray}

\begin{figure}[htb!]
\begin{minipage}[t]{0.5\textwidth}
        \centering
        \includegraphics[width=\textwidth]{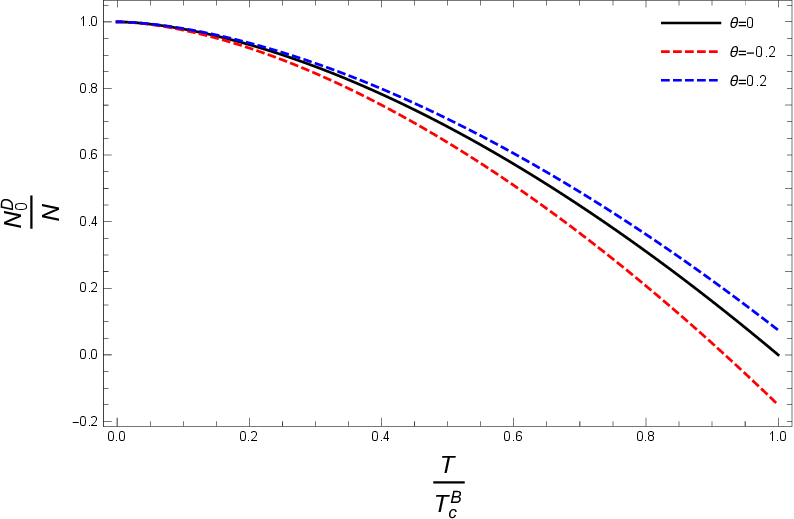}

       \subcaption{ $ \eta=3$}\label{fig:pc2}
   \end{minipage}%
\begin{minipage}[t]{0.5\textwidth}
        \centering
        \includegraphics[width=\textwidth]{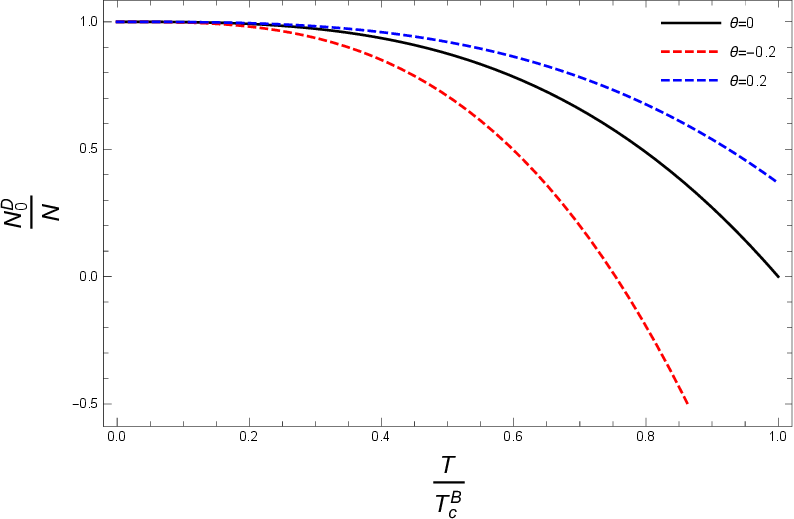}
         \subcaption{ $\eta=1$.}\label{fig:pd2}
   \end{minipage}
\caption{The ground state population versus  $\frac{T}{T_{c}^{B}}$ for different values of the Wigner parameter.} \label{fig4}
\end{figure}

In Figure \ref{fig4}, we display the condensate fraction versus the normalized temperature. To visualize the effect of the Dunkl formalism, we consider three different values for the Wigner parameter. We note that the characteristic effect is the same as in the one-dimensional case.   

\section{Conclusion}

Recent studies that subject the Dunkl formalism bring new insight into the physical problems in two-folds: the derivation of the odd and even parity-based solutions and determination of the free parameters that can be used to provide a better fit between experimental and theoretical results. In this work, we consider an ideal Bose gas system trapped by a one and two-dimensional power-law potential in the Dunkl formalism and investigate the Dunkl-BEC temperature and the ground state population by substituting the Dunkl derivative with the ordinary derivative.  Basically, the Dunkl formalism leads to a modification of the Bose function. The latter function, called the Dunkl-Bose function, consequently affects the BEC temperature and ground state population. We find that these impacts can be similar for the one and two-dimensional traps, for example for the ground state population, and the characteristic behavior of the critical temperature. However, there are still some differences; for example, the constraint on the trap potential parameter in one dimension is not seen in two dimensions.

\section*{Acknowledgments}

This work is supported by the Ministry of Higher Education and Scientific Research, Algeria under the code: PRFU:B00L02UN020120220002. 

\section*{Data Availability Statements}

The authors declare that the data supporting the findings of this study are
available within the article.

\end{document}